\documentstyle{article}
\begin{document}
\begin{center}

{\large\bf Extraction of nuclear matter properties from nuclear
masses by a model of equation of state\\}

{K. C. Chung\footnote{E-mail address: chung@uerj.br}, C. S. Wang and
A. J. Santiago\\}

{\it Instituto de F\'\i sica, Universidade do Estado do Rio de Janeiro,\\
  Rio de Janeiro-RJ 20559-900, Brazil}

\end{center}

\noindent{{\small \bf Abstract}\\

The extraction of nuclear matter properties from measured nuclear
masses is investigated in the energy density functional formalism
of nuclei. It is shown that the volume energy $a_1$ and the nuclear
incompressibility $K_0$ depend essentially on $\mu_n
N+\overline{\mu}_p Z-2E_N$, whereas the symmetry energy $J$ and the
density symmetry coefficient $L$ as well as symmetry
incompressibility $K_s$ depend essentially on
$\mu_n-\overline{\mu}_p$, where $\overline{\mu}_p=\mu_p-\partial
E_C/\partial Z$, $\mu_n$ and $\mu_p$ are the neutron and proton
chemical potentials respectively, $E_N$ the nuclear energy, and
$E_C$ the Coulomb energy. The obtained symmetry energy is
$J=28.5MeV$, while other coefficients are uncertain within
ranges depending on the model of nuclear equation of
state.}\\


\section{Introduction}

The main measured quantities which are used to extract the
properties of standard nuclear matter are the nuclear mass $M$ and
nuclear isoscalar giant monopole resonance energy $E_M$. By
nuclear matter we mean the uncharged nucleon system distributed
uniformly in the space, and by standard nuclear matter we mean the
groundstate nuclear matter with equal neutron and proton numbers.
By properties of standard nuclear matter we mean the volume energy
coefficient $a_1$, the symmetry energy coefficient $J$, the
incompressibility coefficient $K_0$, the density symmetry
coefficient $L$, and the symmetry incompressibility coefficient
$K_s$. Nowadays the most interesting quantity for heavy ion
collision studies as well as for supernova explosion calculations
is the nuclear incompressibility $K_0$.

In the earlier stage of the nuclear matter study, the nuclear
incompressibility $K_0$ was determined in a compressible model of
nuclei, e.g., the droplet model\cite{Myers69}, from nuclear masses
$M$. A previous calculation by this approach based on Seyler-Blanchard
interaction gave $K_0=306MeV$\cite{Groote}, and a recent investigation
by using the droplet model based on the generalized Seyler-Blanchard
interaction yielded $K_0=234MeV$\cite{Myers96}. On the other hand,
in a collective model of the nuclear breathing mode, e.g., the model
based on the scaling assumption\cite{Blaizot76}, $K_0$ can be
extracted from nuclear monopole resonance energies $E_M$. An
earlier attempt in this line based on Skyrme interaction gave
$K_0=210MeV$\cite{Treiner}. However, a later study claimed that
more accurate data of monopole resonance energy lead to a higher
value $K_0\approx 300MeV$\cite{Sharma}.

The staggering of the values of $K_0$ stimulated many works on this
topic\cite{Blaizot95}-\cite{Chossy}. The purpose of this paper is to 
study the problem of extracting the nuclear incompressibility $K_0$
from nuclear masses $M$ as directly as possible. In addition,
 the extraction of the volume
energy $a_1$, the symmetry energy $J$, the density symmetry
coefficient $L$, and the symmetry incompressibility $K_s$ is
discussed simultaneously. In Sec.2 the theory used in the present
work is formulated. A model of nuclear equation of state is
presented in Sec.3, while the approximation by leptodermous
expansion is given in Sec.4. In Sec.5 a short discussion and
summary are addressed.

\section{Theory}
The total energy $E(A, Z)$ of a nucleus with nucleon number $A$
and proton number $Z$ can be written as
\begin{equation}
 \label{EAZ}
 E(A,Z)=E_N+E_C+E_{res},
\end{equation}
where $E_N$ is the nuclear energy, $E_C$ the Coulomb energy, and
$E_{res}$ the residual energy which includes the shell correction,
the even-odd energy and the congruence energy\cite{Myers96}. The
nuclear energy $E_N$ can be expressed as
\begin{equation}
 \label{EN} E_N=\int d^3r{\cal E}_N,
\end{equation}
where the nuclear energy density functional ${\cal E}_N$ can be
written with enough generality as
\begin{equation}
 \label{calEN}
 {\cal E}_N=\rho({\bf r})e(\rho,\delta)+\frac{1}{2}Q_1(\nabla\rho)^2
  +Q_2\Big[(\nabla\rho_n)^2+(\nabla\rho_p)^2\Big].
\end{equation}
In above equation, $\rho_n$ and $\rho_p$ are the neutron and
proton densities, respectively, $\rho=\rho_n+\rho_p$ the nucleon
density,
\begin{equation}
 \label{delta} \delta=\frac{\rho_n-\rho_p}{\rho} ,
\end{equation}
the relative neutron excess, $e(\rho,\delta)$ the nuclear equation
of state, and $Q_1, Q_2$ the model parameters related to the
finite size and surface effects of nuclei.

The Euler-Lagrange equations obtained by the minimization of
$E(A,Z)$ are
\begin{equation}
 \label{EL1}
 e+\rho\frac{\partial e}{\partial \rho}+(1-\delta)\frac{\partial e}
 {\partial\delta}-Q_1\nabla^2\rho-2Q_2\nabla^2\rho_n-\mu_n=0,
\end{equation}
\begin{equation}
 \label{EL2}
 e+\rho\frac{\partial e}{\partial \rho}-(1+\delta)\frac{\partial e}
 {\partial\delta}-Q_1\nabla^2\rho-2Q_2\nabla^2\rho_p-\overline{\mu}_p
 =0,
\end{equation}
\begin{equation}
\label{mup}
 \overline{\mu}_p=\mu_p-\frac{\partial E_C}{\partial Z},
\end{equation}
where $\mu_n$ and $\mu_p$ are the neutron and proton chemical
potentials, respectively. Eqs.(\ref{EL1}) and (\ref{EL2}) can
be reduced to
\begin{equation}
\label{EL1a}
 e+\rho\frac{\partial e}{\partial\rho}
  -\delta\frac{\partial e}{\partial\delta}-Q\nabla^2\rho
  =\frac{1}{2}(\mu_n+\overline{\mu}_p),
\end{equation}
\begin{equation}
\label{EL2a}
 \frac{\partial e}{\partial\delta}-Q_2\nabla^2(\rho\delta)
  =\frac{1}{2}(\mu_n-\overline{\mu}_p),
\end{equation}
where $Q=Q_1+Q_2$. By substituting $\partial e/\partial\delta$ solved
from Eq.(\ref{EL2a}) into Eq.(\ref{EL1a}), it results
\begin{equation}
\label{EL1b}
 e+\rho\frac{\partial e}{\partial\rho}-Q\nabla^2\rho
  -Q_2\delta\nabla^2(\rho\delta)
 =\frac{1}{2}\big[(1+\delta)\mu_n+(1-\delta)\overline{\mu}_p\big].
\end{equation}
It is worthwhile to note that, if the nuclear surface and the Coulomb
energies are omitted and the equilibrium condition
$\partial e/\partial\rho=0$ is assumed, the above equation is reduced
to
\begin{equation}
 \label{gHVH} e=\frac{1}{2}\big[(1+\delta)\mu_n+(1-\delta)\mu_p\big].
\end{equation}
This is the so-called generalized Hugenholtz-Van Hove
theorem\cite{Satpathy} which is used successfully in nuclear
mass data fit\cite{Nayak}.

By integrating Eqs.(\ref{EL2a}) and (\ref{EL1b}) over the nucleon
density, the following relationships can be obtained:
\begin{equation}
 \label{EQ2}
 \int d^3 r \rho\Big[\frac{\partial e}{\partial\delta}
  -Q_2\nabla^2(\rho\delta)\Big]=\frac{1}{2}(\mu_n-\overline{\mu}_p)A,
\end{equation}
\begin{equation}
 \label{EQ1}
 \int d^3 r \rho\Big[-e+\rho\frac{\partial e}{\partial \rho}\Big]
  =\mu_nN+\overline{\mu}_pZ-2E_N,
\end{equation}
which are the basic equations used in the present work. In
obtaining Eq.(\ref{EQ1}), Eqs.(\ref{EN}) and (\ref{calEN})
are used.

The nuclear energy $E_N$ can be calculated by the following formula
in $MeV$\cite{Myers96}:
\begin{equation}
 \label{ENexp} E_N=\Delta M-8.071431\,N-7.289034\,Z+0.00001433\,Z^{2.39}
  -E_C-E_{res},
\end{equation}
where the measured mass excess $\Delta M$ can be taken from the
table given in Ref.\cite{Moller}, the residual energy can be
calculated by the empirical formulas of Ref.\cite{Moller}, while the
Coulomb energy can be determined by the following
formula \cite{Hasse}:
\begin{equation}
 \label{EC}
 E_C=\frac{3}{5}\frac{e^2Z^2}{R}\Big[1-\frac{5}{2}\Big(\frac{b}{R}\Big)^2
     +3.0216\Big(\frac{b}{R}\Big)^3+\Big(\frac{b}{R}\Big)^4\Big].
\end{equation}
In above equation, $b$ is the S\"ussmann width which is related to
the surface diffuseness $d$ as $d=\sqrt{3}\,b/\pi$, and
$R=r_0A^{1/3}$, $r_0$ the nuclear radius constant. The Coulomb
exchange energy is not included in this equation, since we want to
make a comparison with the Myers-Swiatecki's Thomas-Fermi
calculation\cite{Myers96} where this contribution is not
considered. However, the inclusion of the Coulomb exchange energy
in the present scheme is easy and we have checked that it gives no
significant change in the numerical result.

The chemical potentials $\mu_n$ and $\mu_p$ can be obtained from
nuclear masses by the following formulas\cite{Nayak}:
\begin{equation}
 \mu_n=\frac 12\big[E(A+1,Z)-E(A-1,Z)\big],
\end{equation}
\begin{equation}
 \mu_p=\frac 12\big[E(A+1,Z+1)-E(A-1,Z-1)\big].
\end{equation}
Similarly, $\partial E_C/\partial Z$ can be calculated as
\begin{equation}
 \label{pECpZ}
 \frac{\partial E_C}{\partial Z}
  =\frac 12\big[E_C(A+1,Z+1)-E_C(A-1,Z-1)\big].
\end{equation}

Thus, the righthand sides of Eqs.(\ref{EQ2}) and (\ref{EQ1}) are
known from the measured mass excess $\Delta M$ and the calculated
Coulomb energy $E_C$ as well as the empirical residual energy
$E_{res}$. Therefore, these two equations can be used to determine
the equation of state and thus the nuclear matter properties if
the nucleon density $\rho({\bf r})$ is known also from
experiment.

The $\mu_n-\mu_p$ obtained by this way is shown in Fig.1 versus
the nuclear asymmetry $I=(N-Z)/A$, while $\mu_n-\overline{\mu}_p$
versus $I$ is shown in Fig.2. It can be seen that there is an
approximate linear correlation between $\mu_n-\overline{\mu}_p$
and $I$, but the correlation between $\mu_n-\mu_p$ and $I$ is not
so simple. This correlation will be explained by the leptodermous
expansion given in Sec.4. In addition, Fig.3 shows the data of
$(\mu_nN+\overline{\mu}_pZ-2E_N)/A$ obtained by this way versus
$A^{-1/3}$. The total number of masses is $1654$, while the total
number of $\mu_n-\overline{\mu}_p$ and
$(\mu_nN+\overline{\mu}_pZ-2E_N)/A$ data obtained from these
masses is $1140$ each one.

\section{Model of nuclear equation of state}

It is shown that for a variety of nuclear interactions the
nuclear equation of state can be written with enough generality as
\begin{equation}
\label{EoS}
 e(\rho, \delta)=T\Big[D_2(\delta)\Big(\frac\rho\rho_0\Big)^{2/3}
                -D_3(\delta)\Big(\frac\rho\rho_0\Big)^{3/3}
       +D_\gamma(\delta)\Big(\frac\rho\rho_0\Big)^{\gamma/3}\Big],
\end{equation}
where $\rho_0=3/4\pi r_0^3$ is the standard nuclear matter
density, $T$ an appropriate constant with dimension of energy such
that $D_2(\delta), D_3(\delta)$ and $D_\gamma(\delta)$ are
dimensionless, and $\gamma$ a model parameter. It is convenient to
choose $T$ as the Fermi energy of the standard nuclear matter,
$T=(\hbar^2/2m)(3\pi^2\rho_0/2)^{2/3}$ (See more details in Ref.
\cite{CWS}.

For standard nuclear matter, we have $\partial
e/\partial\rho|_0=0$ at $\rho=\rho_0$ and $\delta=0$, the
following relationship among $D_2(0)$, $D_3(0)$, $D_\gamma(0)$ and
$\gamma$ should be satisfied:
\begin{equation}
 \label{stable} 2D_2(0)-3D_3(0)+\gamma D_\gamma(0)=0.
\end{equation}
The following formulas of nuclear matter properties can be
obtained:
\begin{equation}
 \label{a1} a_1=-e(\rho_0, 0)
    =-\frac T3\big[D_{20}-(\gamma-3)D_{\gamma 0}\big],
\end{equation}
\begin{equation}
 \label{K0} K_0=9\rho_0^2\frac{\partial^2e}{\partial\rho^2}|_0
    =T\big[-2D_{20}+\gamma(\gamma-3)D_{\gamma 0}\big],
\end{equation}
\begin{equation}
 \label{J} J=\frac12\frac{\partial^2e}{\partial\delta^2}|_0
            =T\big[D_{22}-D_{32}+D_{\gamma 2}\big],
\end{equation}
\begin{equation}
 \label{L} L=\frac32\rho_0
   \frac{\partial^3e}{\partial\rho\partial\delta^2}|_0
  =T\big[2D_{22}-3D_{32}+\gamma D_{\gamma 2}\big],
\end{equation}
\begin{equation}
 \label{Ks} K_s=\frac92\rho_0^2
     \frac{\partial^4e}{\partial\rho^2\partial\delta^2}|_0
    =T\big[-2D_{22}+\gamma(\gamma-3)D_{\gamma 2}\big],
\end{equation}
where
\begin{equation}
 D_{i0}=D_i(0),\,\,\,\,
 D_{i2}=\frac 12\frac{\partial^2D_i}{\partial\delta^2}|_0,
 \,\,\,\, i=2, 3, \gamma,
\end{equation}
and the relation (\ref{stable}) is used in obtaining
Eqs.(\ref{a1}) and (\ref{K0}).

 Since
the nuclear interaction is symmetric in proton and neutron,
$D_i(\delta)$ are even functions of $\delta$. The specific
discussion for these dependences is beyond the scope of this paper
and will be given in a forthcoming paper.
Hence, the coefficients
$D_i(\delta)$ can be written approximately as linear function of
$\delta^2$ for the integrals involved in Eqs. (\ref{EQ2}) and
(\ref{EQ1}). In addition, for stable nuclei, where $\delta$ is small,
we have: 
\begin{equation}
 \label{Di}
 D_i(\delta)=D_{i0}+D_{i2}\delta^2,\,\,\, i=2, 3, \gamma.
\end{equation}
Due to the equilibrium condition (\ref{stable}), there are only $2$
independent coefficients among $3$ $D_{i0}$, they can be chosen as
$D_{20}$ and $D_{\gamma0}$. Therefore, there are $5$ independent
coefficients within the above approximation (\ref{Di}): $D_{20}$,
$D_{\gamma0}$, $D_{22}$, $D_{32}$ and $D_{\gamma2}$.
Correspondingly, all of the $5$ nuclear matter properties $a_1$,
$K_0$, $J$, $L$ and $K_s$ are independent each other.

Eqs.(\ref{EQ2}) and (\ref{EQ1}) can be fitted to nuclear masses,
by using $D_{20}$, $D_{\gamma0}$, $D_{22}$, $D_{32}$ and
$D_{\gamma2}$ as free adjustable parameters in the equation of
state, while the following two-parameter Fermi distribution is
assumed for nucleon distribution $\rho({\bf r})$:
\begin{equation}
 \label{Fermi} \rho({\bf r})=\frac{\rho_A}{1+e^{\frac{r-C}d}},
\end{equation}
where
\begin{equation}
 C=R\Big[1-\frac13\Big(\frac{\pi d}R\Big)^2
   +O\Big(\Big(\frac dR \Big)^6\Big)\Big]
\end{equation}
is the nuclear half density radius\cite{Hasse}. In the numerical
calculation, $b=1.0 fm$ and $r_0=1.14 fm$ are
used\cite{Myers96}\cite{Buchinger}.

In addition, the following approximations are assumed:
\begin{equation}
 \label{alphai}
 \int d^3r\rho(r)\delta\Big(\frac{\rho}{\rho_0}\Big)^{i/3}
  \approx I\int d^3r\rho(r)\Big(\frac{\rho}{\rho_0}\Big)^{i/3},
  \,\,\,\,i=2,3,\gamma,
\end{equation}
\begin{equation}
 \label{betai}
 \int d^3r\rho(r)\delta^2\Big(\frac{\rho}{\rho_0}\Big)^{i/3}
  \approx \xi_iI^2\int d^3r\rho(r)
  \Big(\frac{\rho}{\rho_0}\Big)^{i/3},\,\,\,\,i=2,3,\gamma,
\end{equation}
where the $\xi_i$'s are adjustable parameters to be fitted to measured data.
These approximations are reasonable, because $\rho/\rho_0\approx
1$ in the core and Eq.(\ref{alphai}) is exact for $\rho/\rho_0=1$.
Furthermore, the surface region gives negligible contribution to
the integral. A study based on an approximation to
Eq.(\ref{delta}), using two-parameter Fermi distribution for
$\rho_n$ and $\rho_p$, shows that both Eqs. (\ref{alphai}) and
(\ref{betai}) are good approximation.

On the other hand, for the integral involving the Laplacian of
$\rho\delta$ in Eq.(\ref{EQ2}), the following approximation is
employed:
\begin{equation}
 \label{nabla}
 \int d^3r\rho\nabla^2(\rho\delta)\approx -\frac{\rho_Ct}{4R^2}A,
\end{equation}
where $\rho_C=\rho(0)$, $t=(C_n-C_p)/d$, $C_n$ and $C_p$ are the
neutron and proton half density radii respectively. It can be seen
that this term is much smaller than other term in
Eq.(\ref{EQ2}), $t$ can be assumed approximately as an averaged
constant in the data fit.

Numerically, $D_{22}$, $D_{32}$, and $D_{\gamma 2}$ can be
determined by the data fit of Eq. (\ref{EQ2}), and then the
variables $J$, $L$, and $K_s$ can be calculated from these fitted
parameters by Eqs.(\ref{J})-(\ref{Ks}). On the other hand,
$D_{20}$ and $D_{\gamma 0}$ can be determined by the data fit of
Eq.(\ref{EQ1}), and then the variables $a_1$ and $K_0$ can be
calculated from these fitted parameters by Eqs.(\ref{a1}) and
(\ref{K0}). Therefore, the volume energy $a_1$ and nuclear
incompressibility $K_0$ depend essentially on $\mu_n
N+\overline{\mu}_p Z-2E_N$, while the symmetry energy $J$, the
density symmetry coefficient $L$ and the symmetry
incompressibility $K_s$ depend essentially on
$\mu_n-\overline{\mu}_p$. In this way, the present data fit is
separated into two steps, $a_1$ and $K_0$ are extracted from data
of $\mu_n N+\overline{\mu}_p Z-2E_N$, while $J$, $L$, and $K_s$
are extracted from data of $\mu_n-\overline{\mu}_p$.

The result obtained by this data fit depends on the choice of
$\gamma$. N$^{\underline{\rm o}}$ 1 in Table 1 gives the result
with $\gamma=5$. For comparison, the results obtained by this data
fit corresponding to Myers-Swiatecki equation of
state\cite{Myers96} and Tondeur equation of state\cite{Tondeur}
are shown as N$^{\underline{\rm o}}$ 2 and N$^{\underline{\rm o}}$
3 in Table 1 respectively. For Myers-Swiatecki equation of state,
we have $\gamma=5$, and the following restriction can be derived:
\begin{equation}
 \label{D52D22}
  -\frac{6(3D_{22}-1)}{5D_{20}-3}+\frac95\frac{D_{52}}{D_{50}}=2.
\end{equation}
For Tondeur equation of state, in addition to $\gamma=4$, we have
$D_{20}=3/5$ and $D_{32}=D_{\gamma2}=0$ which gives $L=2J$ and
$K_s=-L$. The original values given in Refs.\cite{Myers96} and
\cite{Tondeur} are shown in parentheses of N$^{\underline{\rm o}}$
2 and N$^{\underline{\rm o}}$ 3 in Table 1 respectively. Note the
difference between the present results and the original ones. That
difference can be understood, since the present data fit is
separated into two independent steps which is different from what
is done in Refs.\cite{Myers96} and \cite{Tondeur}, in addition to
some approximation used in the present data fit. 

Fig.4 plots the coefficients $a_1$(circles), $J$(full dots), and $L$(circled
crosses) versus $\gamma$. As $\gamma$ increases from $3.1$ to
$25$, the symmetry energy $J$ increases from $28.50$ to
$28.52MeV$, the volume energy $a_1$ increases from $15.86$ to
$17.47MeV$, while the density symmetry coefficient $L$ decreases
from $61.76$ to $50.97MeV$. Similarly, Fig.5 shows $K_0$ and
$K_s$ versus $\gamma$. As $\gamma$ increases from $3.1$ to $25$,
the nuclear incompressibility $K_0$ increases from $199.2$ to
$673.9MeV$, and the symmetry incompressibility $K_s$ decreases
from $-119.7$ to $-716.4MeV$.

It is interesting to see what will be obtained if the present
model and approximations (\ref{alphai}) and (\ref{betai}) are
applied to the data fit based on mass formula. In present
formulation, Eq.(\ref{EAZ}) together with Eqs.(\ref{EN}),
(\ref{calEN}) and (\ref{EoS}) is equivalent to the mass formula.
Using the same approximations, the following expression can be
written:
\begin{equation}
 \label{Q0k} \frac 12Q_1(\nabla\rho)^2+Q_2\Big[(\nabla\rho_n)^2
 +(\nabla\rho_p)^2\Big]
 \approx \frac 12(Q'+Q_2'\delta^2)(\nabla\rho)^2,
\end{equation}
where $Q'=Q$ and $Q_2'=Q_2$ when $(\nabla\delta)^2$ and
$\nabla\rho\cdot\nabla\delta$ can be neglected, and are the
parameters to be fitted to $E_N$ data. In this case, the
independent parameters to be fitted are: $D_{20}$, $D_{\gamma 0}$,
$\xi_2D_{22}$, $\xi_3D_{32}$, $\xi_\gamma D_{\gamma 2}$, $Q'$, and
$\xi_2'Q_2'$, where $\xi_2'$ is a parameter introduced in an
approximation similar to Eq.(\ref{betai}). With the fitted values
of these parameters, the total energy $E(A,Z)$ and thus the
nuclear mass can be calculated for given $A$ and $Z$. In the same
time, the nuclear volume energy $a_1$ and incompressibility $K_0$
can be calculated also from the fitted $D_{20}$ and $D_{\gamma
0}$.

The result with $\gamma=5$ is shown as N$^{\underline{\rm o}}$ 4
in Table 1. In this data fit, the total of $1654$ $E_N$ data,
which are the same data used in the data fit based on Thomas-Fermi
calculation with a standard deviation $0.655MeV$\cite{Myers96},
are fitted. It is interesting to note that, when $\gamma$
increases from $3.1$ to $25$, $a_1$ varies from $16.034$ to
$16.042MeV$ with a minimum $16.028MeV$ at $\gamma=11$, $K_0$
increases monotonically from $181$ to $647MeV$.

Therefore, neither in the traditional data fit to nuclear energies
$E_N$ nor in the present data fit to both $\mu_n-\overline\mu_p$
and $\mu_nN+\overline\mu_pZ-2E_N$, the nuclear incompressibility
$K_0$ cannot be determined from nuclear masses alone without
additional assumption on the model of equation of state. In this
aspect, we can make a model assumption on $\gamma$, as given in
the Seyler-Blanchard type interaction\cite{Myers96}, or make a
specific choice of $\gamma$, as given in the Tondeur
interaction\cite{Tondeur} or the Skyrme interaction\cite{Brack}.
In addition, we can make some approximation on the equation of
state, as the leptodermous expansion which will be given in the
next Section.

\section{Leptodermous Expansion}

For the nucleon distribution of finite nuclei, there is an
approximate flat plateau in the core and a steep decrease in the
surface region. In this leptodermous distribution of nucleons
$\rho({\bf r})$, the surface region gives a very small
contribution to the integrals on the lefthand side of
Eqs.(\ref{EQ2}) and (\ref{EQ1}). Therefore, the following
expansion of equation of state is expected to be good enough for
these integrals:
\begin{equation}\label{erhodel}
 e(\rho,\delta)=-a_1+\frac 1{18}\big(K_0+K_s\delta^2\big)
 \Big(\frac{\rho-\rho_0}{\rho_0}\Big)^2+\Big[J+\frac L3
 \Big(\frac{\rho-\rho_0}{\rho_0}\Big)\Big]\delta^2.
\end{equation}
In addition, the Fermi distribution (\ref{Fermi}) is a good
approximation for the nucleon density.

By using the leptodermous expansion (\ref{erhodel}) for lefthand
side of Eqs.(\ref{EQ2}) and (\ref{EQ1}), and assuming the Fermi
distribution (\ref{Fermi}) for nucleon density $\rho({\bf r})$, we
can obtain the following approximate formulas:
\begin{equation}
 \label{LEXP2}
  2JI\Big[1-\frac{Ld}{JR}\Big]=\frac 12(\mu_n-\overline{\mu}_p),
\end{equation}
\begin{equation}
 \label{LEXP1} a_1-\frac{K_0d}{4R}\Big[1-\frac{2d}{3R}\Big]
  -J\Big[1-\frac L{3J}\Big]I^2
  =\frac 1A(\mu_nN+\overline{\mu}_pZ-2E_N).
\end{equation}
In obtaining Eq.(\ref{LEXP2}), the second term in the integrand on
the lefthand side of Eq.(\ref{EQ2}) has been neglected, as its
integral is a small quantity of order of $(d/R)^2$\cite{WCS}(see
Sec.3). Equation (\ref{LEXP1}) is approximately valid for nuclei
with small relative neutron excess $\delta$, as the terms
proportional to $t$ and $t^2$ are neglected in the derivation, and
$t$ is proportional to $\delta$\cite{Myers85}.

Equation (\ref{LEXP2}) can be reduced to
\begin{equation}
 \label{LEXP2a} \mu_n-\overline{\mu}_p\approx 4JI,
\end{equation}
if the term $Ld/JR$ is neglected comparing to $1$. This relation
gives an intuitive geometric picture for the symmetry energy $J$
in Fig.2, i.e., $4J$ is the slope of the curve approximately.
Numerically, the symmetry energy $J$ and density symmetry
coefficient $L$ can be extracted by fitting Eq.(\ref{LEXP2}) to
the $\mu_n-\overline{\mu}_p$ data. This fit to $1140$ data of
nuclei gives $J=28.16 MeV$, $L=70.69 MeV$, and the result is shown
in Fig.6 as $(\mu_n-\overline{\mu}_p)/(1-Ld/JR)$ versus $I$. The
slope of the straight line in this figure is $4J$.

These fitted $J$ and $L$ can be substituted into Eq.(\ref{LEXP1}),
and then the volume energy $a_1$ and the nuclear incompressibility
$K_0$ can be extracted by fitting Eq.(\ref{LEXP1}) to the data of
$(\mu_nN+\overline{\mu}_pZ-2E_N)/A$ with small $|\delta|$. The
fitted result is shown in Fig.7 as
$(\mu_nN+\overline{\mu}_pZ-2E_N)/A+(J-L/3)I^2$ versus $A^{-1/3}$.
It can be seen from Eq.(\ref{LEXP1}) that the extrapolation of the
curve to infinite nuclei will give the volume energy $a_1$, while
the minus slope of the curve on the lefthand side will be
approximately proportional to nuclear incompressibility $K_0$. So
Eq.(\ref{LEXP1}) gives an intuitive geometric picture for $a_1$
and $K_0$ in this plot. This fit to $222$ data of nuclei with
$|\delta|<0.1$ gives $a_1=15.29MeV$ and $K_0=225.7MeV$, as shown
as N$^{\underline{\rm o}}$ 5 in Table \ref{table1}.

\section{Discussion and Summary}

The present way of data fit is similar to what is proposed in
Ref.\cite{Nayak} in two aspects. At first, the data fit is
separated into two steps instead of a unique one. In the present
work, the first step gives $J$, $L$ and $K_s$ when the model of
equation of state is assumed, based on the data of
$\mu_n-\overline{\mu}_p$, while the second step gives $a_1$ and
$K_0$ based on the data of $\mu_nN+\overline\mu_p-2E_N$.
Therefore, it is easy to see how good the fit is in a more direct
and intuitive way. Actually, it can be seen from Figs.2 and 3, or
the corresponding Figs.6 and 7, that the fitted $J$ is more
accurate than the fitted $a_1$ and $K_0$ in the present data fit.
Secondly, what is used in the data fit is not only the nuclear
masses but also the separated energies $\mu_n$ and
$\overline{\mu}_p$(in the present work), or $\mu_p$(in
Ref.\cite{Nayak}). This means that the correlations among nuclear
masses are taken into account in this kind of data fit. Even the
shell effect is not treated consistently in this work as was done
in Ref.\cite{Nayak} but rather is calculated by the empirical
formula of Ref.\cite{Moller}, the present result is still
comparable with Ref.\cite{Nayak}. It is noteworthy for that their
$a_1=16.11MeV$ is very close to what we obtained with $\gamma=5$.

It can be seen from Figs. 4 and 5 as well as Table 1 that there is
almost no model dependence in the symmetry energy $J$, while the
model dependence in the volume energy $a_1$ and density symmetry
coefficient $L$ is weak. However, there is strong model dependence
in the nuclear incompressibility $K_0$ and symmetry
incompressibility $K_s$. In this aspect, the choice of $\gamma$
should be considered as an ingredient of the model. This model
dependence is unescapable whenever $K_0$ can not be measured
directly from the definition $K_0=9\rho^2_0\partial^2
e/\partial\rho^2|_0$. It is worthwhile to note that the value of
$K_0$ obtained from different nuclear measurements and
astrophysical observations are spread over a large range from
$180$ to $800MeV$\cite{Norman}, and the value of $K_s$ is between
$-566\pm1350$ to $34\pm159MeV$ obtained by data fit to
breathing-mode energies\cite{Shlomo} while between $-400MeV$ to
$466MeV$ estimated by various types of nuclear force
calculations\cite{Li}.

This indeterminancy of $K_0$ can be understood from Fig.7. As the
nuclear incompressibility $K_0$ is approximately proportional to
minus slope of the curve on the lefthand side, it is hard to be
determined due to the wide spread of the mass data, even if based on
selected $222$ points instead of all $1654$ points as shown in
Fig.3. This spread of mass data supports reasonably the
understanding that $K_0$ cannot be determined from nuclear masses
alone, and some additional measurement such as breathing-mode
energies is required also. As the model of equation of state is
essential in the formulation of the model of nucleon interaction,
indeterminancy of $K_0$ means that the model of nucleon
interaction can not be determined from nuclear masses alone, and
this is well understood in the field.

The model of equation of state is the same for N$^{\underline{\rm
o}}$ 1 and N$^{\underline{\rm o}}$ 4 in Table 1, with same value
of $\gamma=5$, and the fitted $a_1$ and especially $K_0$ are also
very close each other. This means that the present data fit,
separated into two steps, is consistent with the data fit based on
the nuclear mass formula. Therefore, the present model provides
also an appropriate nuclear mass formula, if it is fitted to
nuclear masses.\\

\noindent{\bf Acknowledgments}\\

We are grateful to Dr.W.D.Myers for the reports and reprints. The
authors would like to acknowledge the support from the Funda\c
c\~ao de Amparo \`a Pesquisa do Estado do Rio de Janeiro (FAPERJ).

\vskip 0.3cm
{\noindent\large\bf Figure Captions\\}

{\noindent\bf Figure 1.} $\mu_n-\mu_p$ obtained from nuclear
masses versus nuclear asymmetry $I=(N-Z)/A$.\\

{\noindent\bf Figure 2.} $\mu_n-\overline{\mu}_p$ versus
nuclear asymmetry $I=(N-Z)/A$. It can be seen that there is an
approximate linear correlation for $\mu_n-\overline{\mu}_p$, but
the correlation between $\mu_n-\mu_p$ and $I$ is not so simple.\\

{\noindent\bf Figure 3.} $(\mu_nN+\overline{\mu}_pZ-2E_N)/A$ versus
$A^{-1/3}$.\\

{\noindent\bf Figure 4.} The coefficients $a_1$(circles), $J$(full dots), and
$L$(circled crosses) versus $\gamma$.\\

{\noindent\bf Figure 5.} $K_0$ and $K_s$ versus $\gamma$.\\

{\noindent\bf Figure 6.} $(\mu_n-\overline{\mu}_p)/(1-Ld/JR)$
versus $I$, where $R=r_0A^{1/3}$, $b=1.0 fm$ and $r_0=1.14 fm$ are
used. The slope of the straight line in this figure is $4J$. The
data fit gives $J=28.16 MeV$, $L=70.69 MeV$.\\

{\noindent\bf Figure 7.}
$(\mu_nN+\overline{\mu}_pZ-2E_N)/A+(J-L/3)I^2$ versus $A^{-1/3}$.
It can be seen from  Eq.(\protect\ref{LEXP1}) that the
intersection of the curve at $A^{-1/3}=0$ gives the volume energy
$a_1$, while the minus slope of the curve on the lefthand side is
approximately proportional to nuclear incompressibility $K_0$. The
data fit gives $a_1=15.29 MeV$ and $K_0=225.7 MeV$.\\

\begin{table}[h]
\caption{The fitted values of $a_1$, $K_0$, $J$, $L$, and $K_s$,
all in $MeV$. N$^{\underline{\rm o}}$ 1 is the result of
Eqs.(\protect\ref{EQ2}) and (\protect\ref{EQ1}) with $\gamma=5$.
N$^{\underline{\rm o}}$ 2 is the result corresponding to
Myers-Swiatecki equation of state, while N$^{\underline{\rm o}}$ 3
is the result corresponding to Tondeur equation of state.
N$^{\underline{\rm o}}$ 4 is the result of Eq.(\protect\ref{EN})
together with Eqs.(\protect\ref{calEN}), (\protect\ref{EoS}),
(\protect\ref{Q0k}), and $\gamma=5$. The values shown in
parentheses in N$^{\underline{\rm o}}$ 2 and N$^{\underline{\rm
o}}$ 3 are given by Refs.\protect\cite{Myers96} and
\protect\cite{Tondeur} respectively. N$^{\underline{\rm o}}$ 5 is
the result of the leptodermous expansion. See the text for
details.} \vspace{15pt}
\begin{tabular}{lcccccc}
\hline
 No. & $a_1$ & $K_0$ & $J$   & $L$   & $K_s$   & $\gamma$ \\
\hline
  1  & 16.10 & 237.9 & 28.50 & 60.33 & -157.9  & 5 \\
  2  & 16.10 & 237.9 & 28.82 & 81.81 &   40.4  & 5 \\
     &(16.24)&(234~~)&(32.65)&(49.9~)&         &   \\
  3  & 16.63 & 243.9 & 27.52 & 55.05 &  -55.0  & 4 \\
   &(~15.978)&(235.8)&(32.12)&       &         &   \\
  4  & 16.03 & 237.8 &       &       &         & 5 \\
  5  & 15.29 & 225.7 & 28.16 & 70.69 &         &   \\
\hline
\end{tabular}
\label{table1}
\end{table}

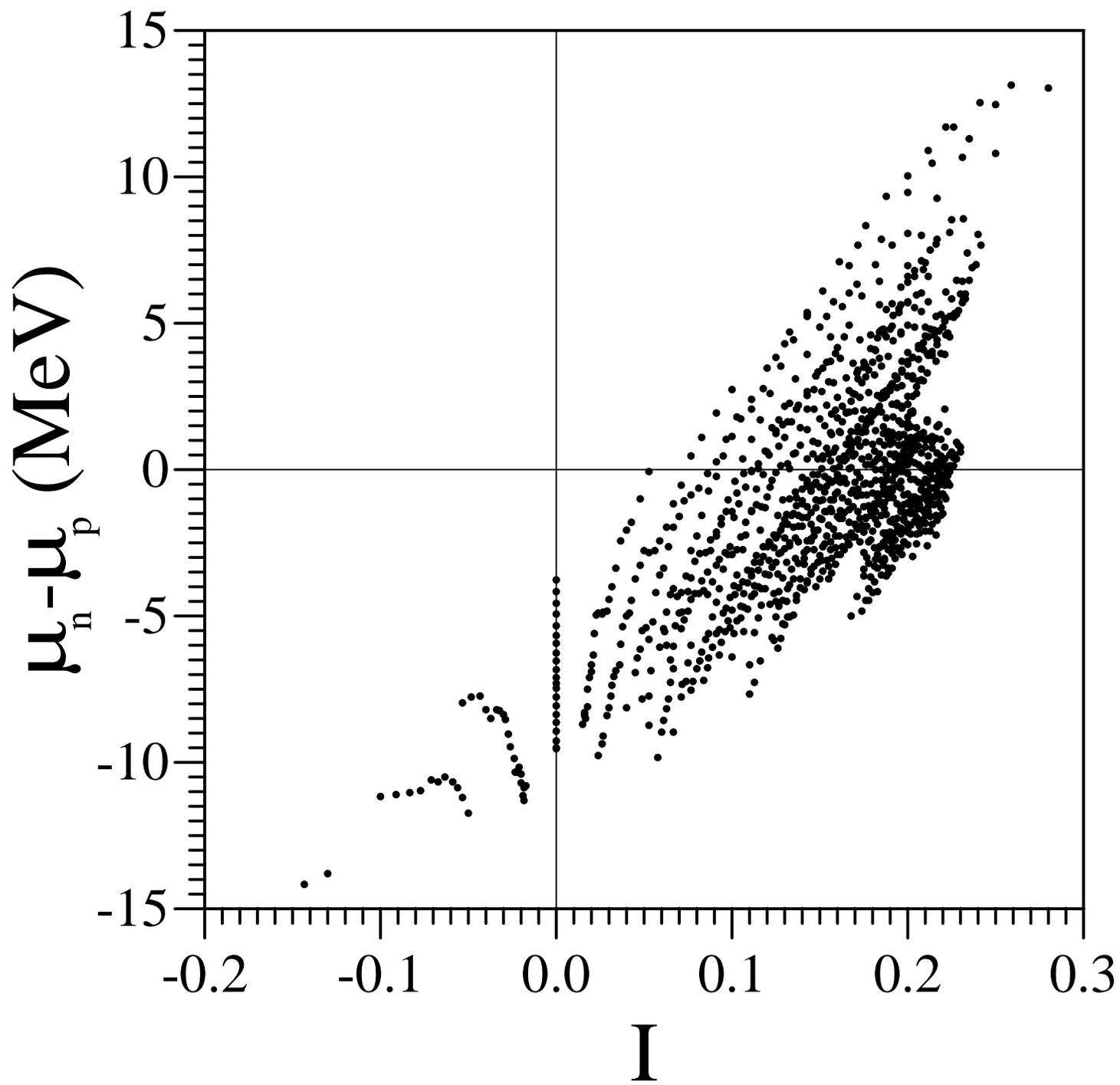
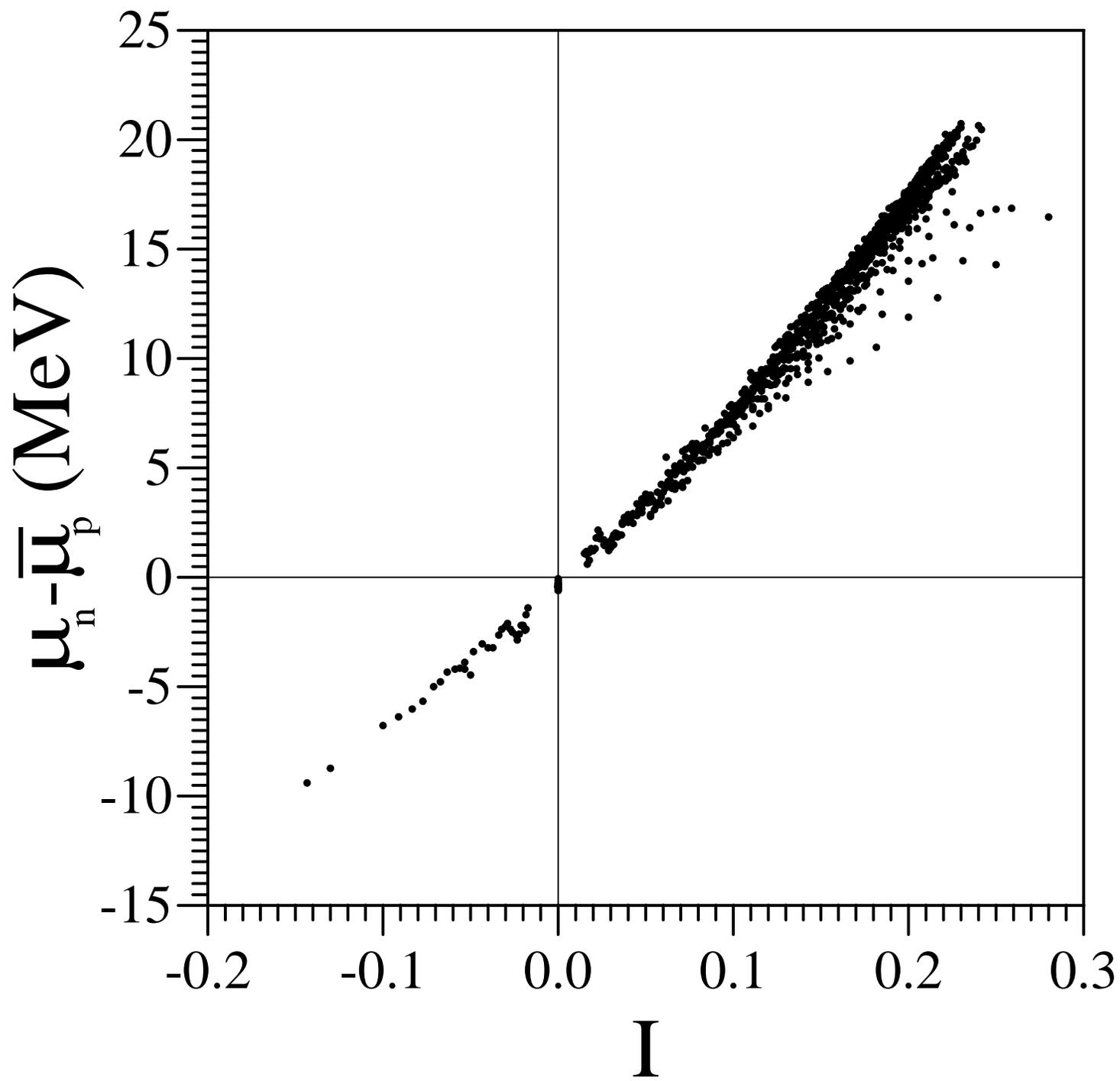
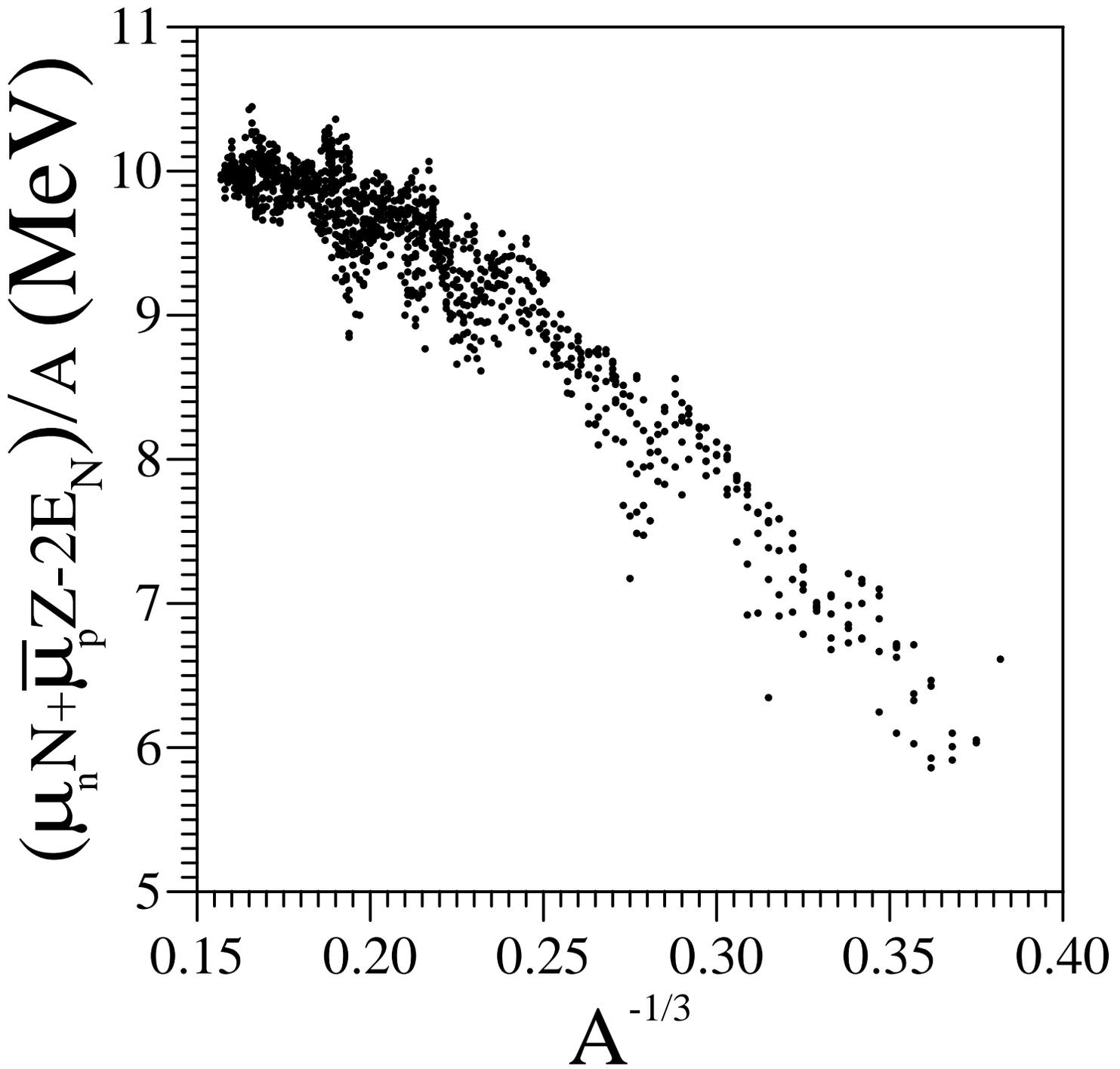
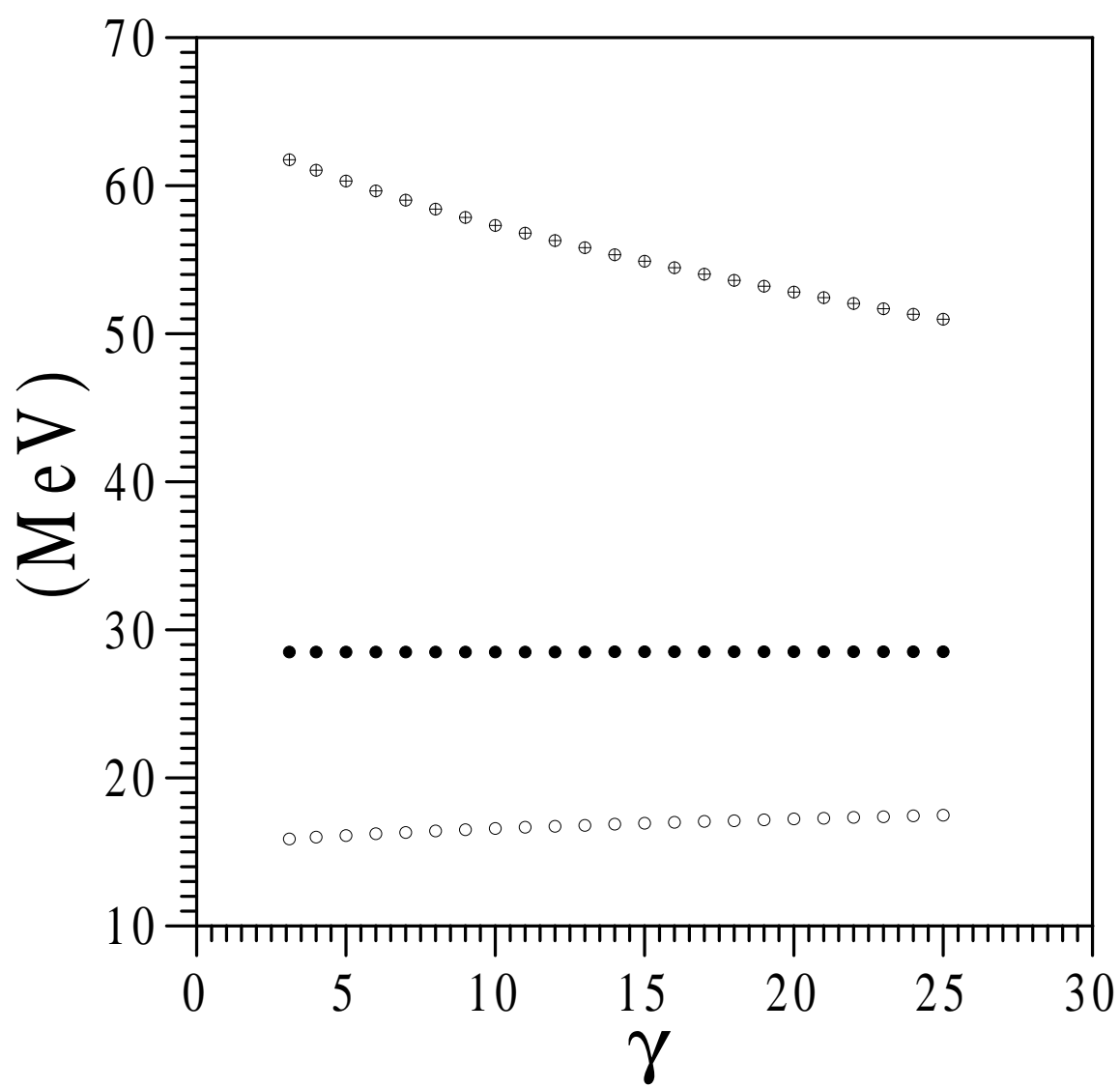
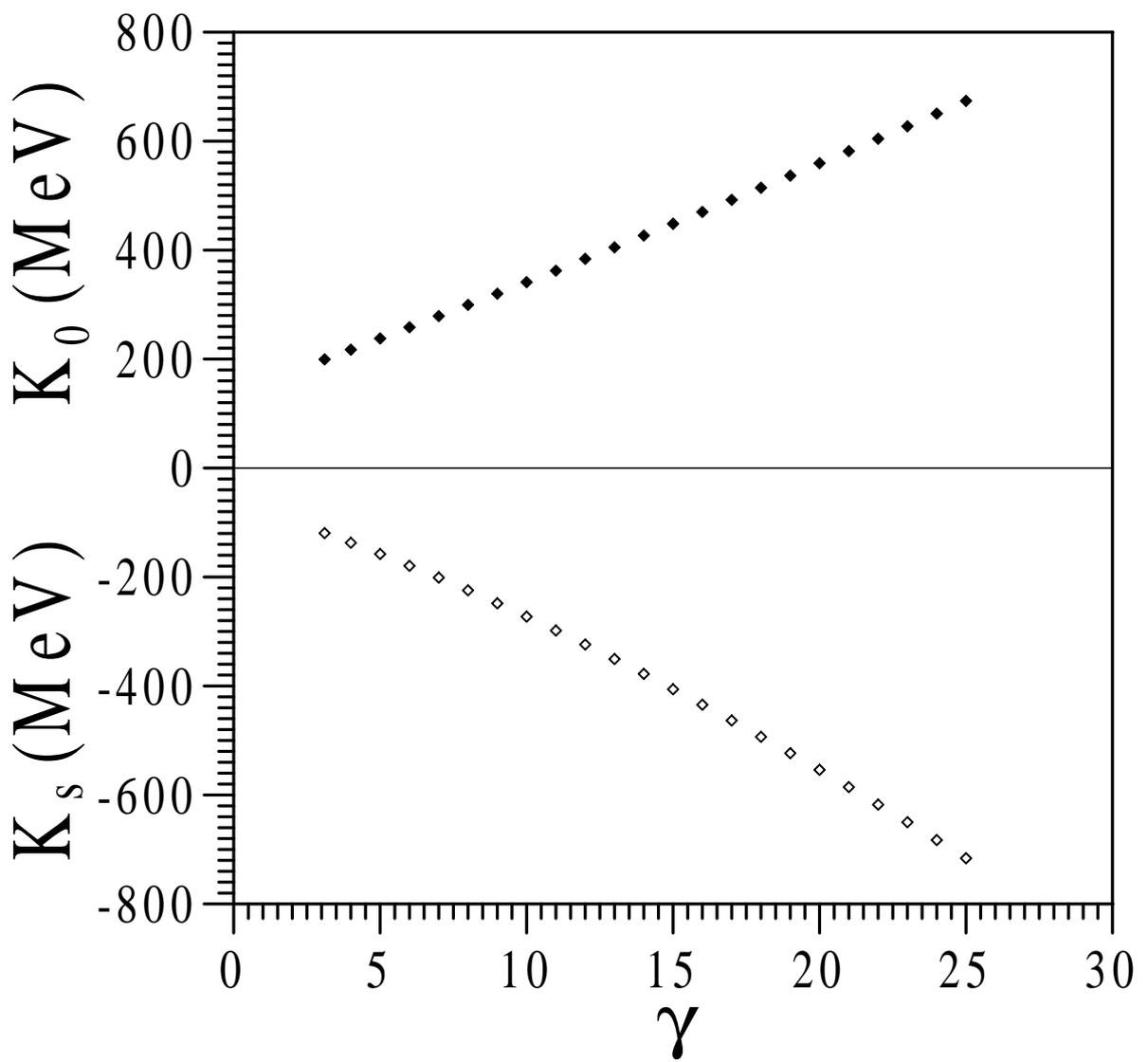
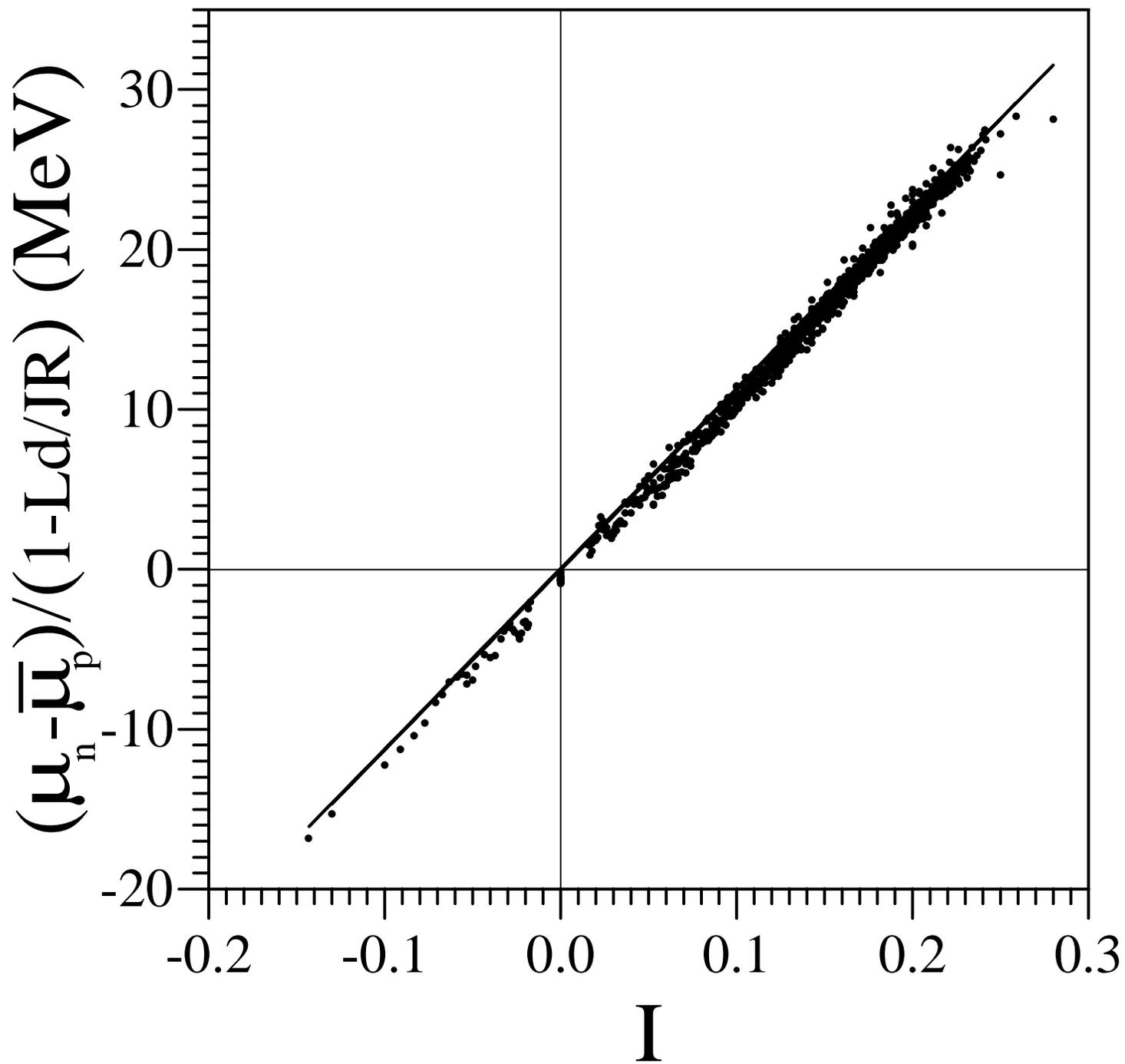
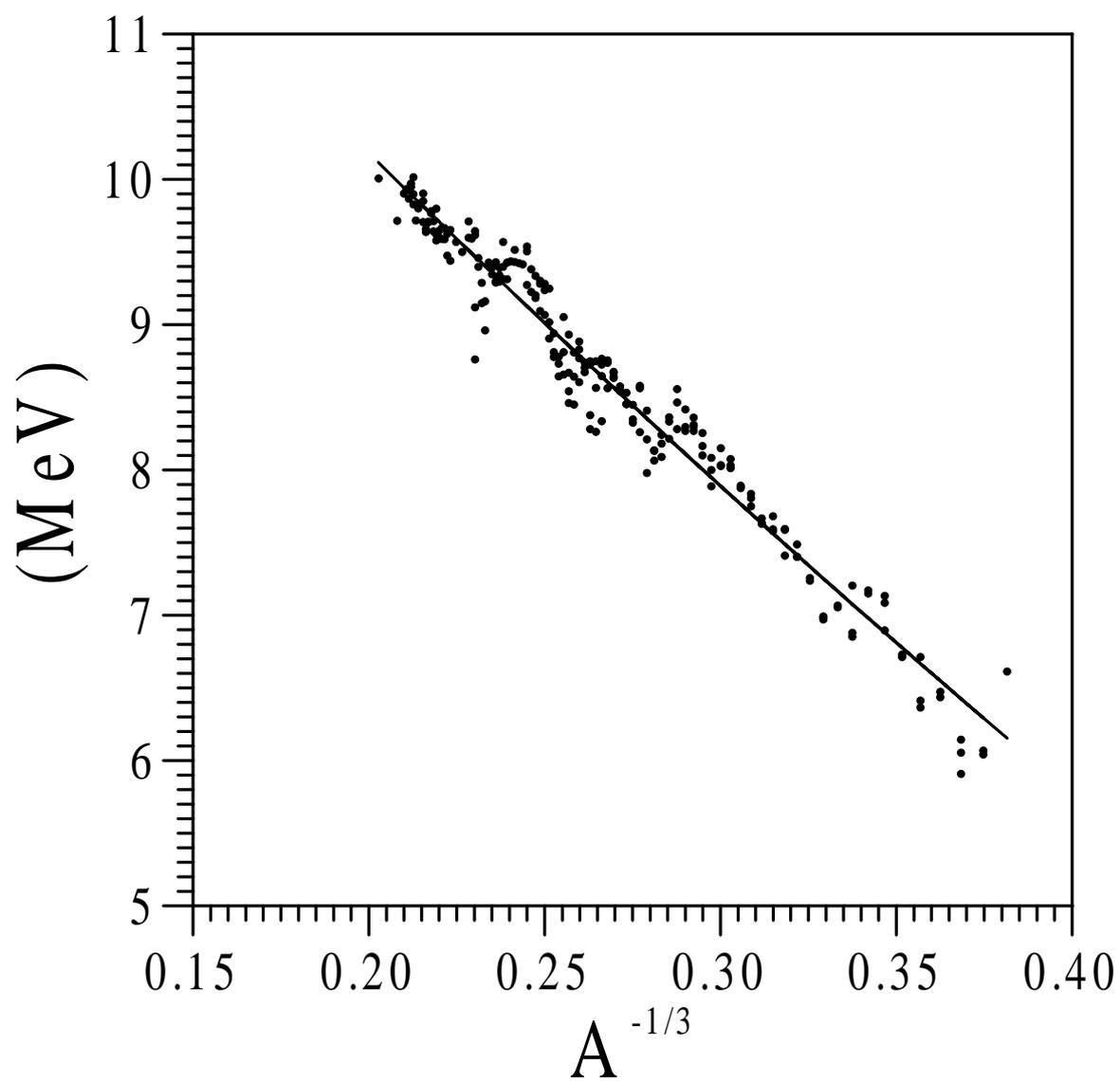
\end{document}